\documentclass[12pt,a4paper]{article}
\usepackage{graphicx,amsmath,amsfonts}
\def \ba{\begin{eqnarray}}\def\ea{\end{eqnarray}}
\def\bc{\begin{center}}\def\ec{\end{center}}
\def\nn{\nonumber\\}
\title{\Large \bf The isospin symmetry  breaking effects in  $K_{e4}$ decays}
\author{\large \bf S.R.Gevorkyan\footnote{On leave of absence from
Yerevan Physics Institute},~~A.N.Sissakian,~~A.V.Tarasov,\\
\bf ~~H.T.Torosyan,~~O.O.Voskresenskaya\footnote{On leave of absence
from Siberian Physical Technical Institute}}
\begin{document}
\maketitle \bc Joint Institute for Nuclear Research, 141980 Dubna,
Russia \ec
\begin{abstract}
 The Fermi-Watson theorem  is generalized to the case of two
coupled channels with different masses and applied to final state interaction
in $K_{e4}$ decays. The impact of  considered effect on the phase of
$\pi\pi$ scattering  is  estimated  and shown that it can be crucial for the scattering lengths
extraction from experimental  data on $K_{e4}$ decays.
\end{abstract}
\section{Introduction}
The $\pi\pi$  scattering at low energies provides a testing ground
for strong interaction study ~\cite{M}. As the free pion targets
cannot be created, the experimental evaluation of $\pi\pi$ scattering
characteristics  is restricted to the study of a dipion system in a
final state of more complicated reactions. One of the most suitable
 for such study processes  are the $K_{e4}$ decays:
 \ba K^\pm \to \pi^+\pi^- e^\pm \nu\ea
 \ba K^\pm\to\pi^0\pi^0 e^\pm \nu \ea
  For many years ~\cite{SH,R} the decay (1) was considered as the
cleanest method to determine the s-wave  isospin zero scattering length
$a_0^0$. At present its value  is predicted by Chiral
Perturbation Theory with  high precision~\cite{CGL} $a_0^0=0.220\pm
0.005$, thus the extraction of this quantity from experimental data
with highest possible accuracy becomes an actual task. The appearance of new  precise
 experimental data~\cite{pis,bb,bat1} requires  the relevant theoretical approaches, taking 
   into account the effects  neglected up to now in comparison
of experimental data  with    theoretical models predictions.\\
The usual method using  in the scattering length  extraction
from decays (1) and (2)  is based on the classical works
~\cite{CM,PT}. The  $K_{e4}$ decay rates are determined by three
axial form factors \footnote{The form factor R is suppressed by a factor
$\frac{m_e^2}{S_e}$ and can't be determined from    $K_{e4}$ decay}
F,G,R and  vector form factor  H. Making the partial-wave expansion of the hadronic
current with respect to the angular momentum of the dipion system
and restricted to s- and p-waves \footnote{As was shown in~\cite{CKS},
the contribution of higher waves are small and can be safely
neglected.} the hadronic  form factors can be written  in the following form:
 \ba
F&=&f_se^{i\delta_s(s)}+f_pe^{i\delta_p(s)}
\cos{\theta_{\pi}}\nn
G&=&g_pe^{i\delta_p(s)};~~~H=h_pe^{i\delta_p(s)}
\ea
Here $s=M_{\pi\pi}^2$ is  the square of  the dipion invariant mass
 and  $\theta_{\pi}$ is the polar angle of pion in the dipion rest frame measured
with respect to the flight direction of dipion in the K- meson rest
frame. The coefficients $f_s,f_p,g_p,h_p$ can be parameterized as a
functions of pions momenta q in dipion rest system  and invariant
mass of lepton pair $s_{e\nu}$ .
It is widely  accepted that the s and p- wave phases $\delta_s $  and  $ \delta_p$ of
dipion system due to Fermi---Watson theorem  \cite{W} coincide  with
the corresponding phase shifts in elastic $\pi\pi$ scattering.Nevertheless this statement
is true if the    isospin symmetry  takes place.   From the other hand
 in the real world   the isospin symmetry breaking effects~\cite{GR,GSTTV,G}
 would play an important part  leading to corrections, which can be  essential in
  scattering length  extraction from  $K_{e4}$ decays\footnote{
  The isospin breaking effects in photoproduction were  considered in ~\cite{An}}\\
Recently, in experiment NA48/2 at CERN~\cite{bat2}  in the $\pi^0\pi^0$
mass distribution from the decay  $K^\pm\to\pi^\pm\pi^0\pi^0$ the
effect of  cusp was observed, which as was pointed by N. Cabibbo
~\cite{C} is the result of isospin symmetry  breaking in final state
$\pi^0\pi^0$ interaction, provided by inelastic $\pi\pi$ reactions
and difference in masses of neutral and charge pions\footnote{The
possibility of  cusp in  $\pi^0\pi^0$ scattering due to different
pion masses in charge exchange reaction $\pi^+\pi^-\to \pi^0\pi^0$
was firstly predicted in ~\cite{MMS}.}.\\
The same effects can take place in $K_{e4}$ decays.
Usually the final state interaction of two pions in $K_{e4}$ decay are
considered using the Fermi-Watson theorem~\cite{W}, which is valid
only in the isospin symmetry limit i.e. at $m_c=m_0$. The main
result of present work is the generalization of   accepted
approach to $K_{e4}$ decays, taking into account the inelastic
processes in the final state and different masses of neutral and
charged pions.
\section{Final state interactions and isospin symmetry breaking}
The s-wave  phase shift  $\delta_s$   has impact only on
hadronic form factor F, whereas the form factors G and H depends only
on p-wave phase shift $\delta_p$.If one confines by
 s and p- waves contributions the charge exchange  process
$\pi^+\pi^-\to \pi^0\pi^0$ and vice versa  are forbidden  for
$\pi\pi$ system in the state with l=1 (p-wave)  due to identity  of neutral
pions. Thus, the inelastic transitions can change only the first  (s-wave) term
in the hadronic form factor F.\\
Keeping this in mind let us denote by $T_c$ the decay amplitude
corresponding   to two charged s-waves pions in  the final state, whereas  the
s-wave amplitude for K meson  decay  to two
neutral pions    is  $T_n$. In one loop approximation of nonperturbative
effective field theory (see e.g. \cite{Gass}) these amplitudes take
the form:
\ba
T_n&=&\tilde{T}_n(1+ik_na_n(s))+ik_ca_x(s)\tilde{T}_c\nn
T_c&=&\tilde{T}_c(1+ik_ca_c(s))+ik_na_x(s)\tilde{T}_n
\ea
Here, $\tilde{T}_c,\tilde{T}_n$ are  so called~\cite{C,CI} ``unperturbed''
amplitudes of decays (1) and (2);
$k_n=\frac{\sqrt{s-4m_0^2}}{2},~~k_c=\frac{\sqrt{s-4m_c^2}}{2}$
are the neutral and charged pions  momenta  in  $\pi^0\pi^0$ and $\pi^+\pi^-$
systems with the same invariant mass $s=M_{\pi\pi}^2$. The  real functions~\cite{C,CI}
 $a_n(s),a_c(s),a_x(s) $  in the isospin symmetry limit ($k_1=k_2=k$)  can be expressed
  through the s-wave  pion-pion "amplitudes" with definite isospin, which  at the
   threshold  coincide with  relevant scattering lengths $a_0^0,a_0^2$:
 \footnote{Our definition of amplitudes coincide
with one adopted  in ~\cite{GORW}  and differs  from accepted in
~\cite{Gass,CI}}:
 \ba
a_n(s)=\frac{a_0(s)+2a_2(s)}{3}; a_c(s)=\frac{2a_0(s)+a_2(s)}{3};
a_x(s)=\frac{\sqrt{2}}{3}(a_0(s)-a_2(s))
\ea
From the other hand these functions are connected with  s-wave  phases with
certain isospin:
\ba
a_0(s)&=&\frac{\tan\delta_0^0(s)}{k};a_2(s)=\frac{\tan\delta_2^0(s)}{k}
\ea
 In the isospin symmetry limit from the rule $\Delta I=1/2$  for semileptonic decays it
follows the simple relation between the ``unperturbed'' amplitudes
$\tilde{T}_c=\sqrt{2}\tilde{T}_n$.  Substituting these relations in   (4)  one gets:
 \ba
T_n&=&\tilde{T}_c(1+ika_0(s))=\tilde{T}_n\sqrt{1+k^2a_0^2(s)}e^{i\delta_0^0(s)}\nn
T_c&=&\tilde{T}_c(1+ika_0(s))=\tilde{T}_c\sqrt{1+k^2a_0^2(s)}e^{i\delta_0^0(s)}
\ea
These equations are nothing else  than  Fermi---Watson theorem for
the pion-pion interaction in final states.\\
But  in the real world, where $m_c\neq m_0$  the
Fermi---Watson theorem in its original form is not valid  and the
two channel problem in this case demands  the special consideration.\\
The considered picture can be generalized~\cite{GTV}  to all orders in $a_j(s)$:
 \ba
T_n&=&\tilde{T}_n(1+ik_nf_n)+ik_cf_x\tilde{T}_c\nn
T_c&=&\tilde{T}_c(1+ik_cf_c)+ik_nf_x\tilde{T}_n
\ea
Here   $f_x,f_c,f_n$ are the amplitudes of the processes
$\pi^+\pi^-\to\pi^0\pi^0; \pi^+\pi^-\to\pi^+\pi^-; \pi^0\pi^0\to\pi^0\pi^0 $
accounting for different masses of charge and neutral pions.These amplitudes can be
expressed through the   real functions $a_x,a_c,a_n$ and relevant  S-matrix elements  :
\ba
S_x&=& 2i\sqrt{k_ck_n}f_x=2i\sqrt{k_ck_n}\frac{a_x(s)}{D}\nn
S_n&=&1+2ik_nf_n=\frac{(1+ik_na_n(s))(1-ik_ca_c(s))-k_nk_ca_x^2(s)}{D}\nn
S_c&=&1+2ik_cf_c=\frac{(1-ik_na_n(s))(1+ik_ca_c(s))-k_nk_ca_x^2(s)}{D}\nn
D&=&(1-ik_na_n(s))(1-ik_ca_c(s))+k_nk_ca_x^2(s)
\ea
In  the isospin symmetry limit ($k_n=k_c=k$) using the expressions (8),(9) and relations
 between S-matrix elements with certain isospin:
\ba
S_c=\frac{2}{3}S_0+\frac{1}{3}S_2;  S_n=\frac{1}{3}S_0+\frac{2}{3}S_2
\ea
after a bit algebra we  obtain:
\ba
S_0&=&\frac{1+ika_0}{1-ika_0}=e^{2i\delta_0}; f_0=\frac{a_0(s)}{1-ika_0(s)}\nn
S_2&=&\frac{1+ika_2}{1-ika_2}=e^{2i\delta_2}; f_2=\frac{a_2(s)}{1-ika_2(s)}
\ea
In the real world  where the  isospin symmetry breaking takes place
the equations (8) can be rewritten   in the following form:
 \ba
 T_n&=&\frac{\tilde{T}_n\sqrt{1+k_c^2(a_c(s)-
\sqrt{2}a_x(s))^2}}{\vert D\vert}e^{i\delta_n}\nn
T_c&=&\frac{\tilde{T}_c\sqrt{1+k_n^2(a_n(s)-\frac{1}{\sqrt{2}}a_x(s))^2}}
{\vert D\vert }e^{i\delta_c}\nn
\delta_n&=&\arctan{\frac{k_na_n(s)+k_ca_c(s)}{1+k_nk_n(a_x^2(s)-a_n(s)a_c(s))}}-\arctan{
k_c(a_c(s)-\sqrt{2}a_x(s))}\nn
\delta_c&=&\arctan{\frac{k_na_n(s)+k_ca_c(s)}
{1+k_nk_c(a_x^2(s)-a_n(s)a_c(s))}}-\arctan
{k_n(a_n(s)-\frac{1}{\sqrt{2}}a_x(s))}\nn
 \ea
In the case of exact isospin symmetry $(m_c=m_0)$ :
\ba
T_n=\frac{\tilde{T}_n}{\sqrt{1+k^2a_0^2(s)}}e^{i\delta_0^0};~~~~
 T_c=\frac{\tilde{T}_c}{\sqrt{1+k^2a_0^2(s)}}e^{i\delta_0^0}
 \ea
which is the  manifestation of Fermi---Watson theorem.\\
 Let us note that unlike the common wisdom, the generalized phases (12) depend
not only on $a_0(s)$, but also on $a_2(s)$.\\
The obtained relations  are valid in the region above the charged pions
production threshold $M_{\pi\pi}=2m_c$. To go under charged pions threshold
in reaction (2) one has to make   the simple substitution $k_n\to
i\kappa$  in the expression for $T_n$ with the result  :
 \ba
T_n&=&\frac{\tilde{T}_ n[1+\kappa
(a_c-\sqrt{2}a_x)]}{D}e^{i\delta_n} \nn \delta_n&=& \arctan
\kappa \left(a_n-\frac{\kappa a_{x}^2}{1+\kappa a_c}\right)
\ea
Thus, as in the case of $K\to 3\pi$  ~\cite{bat2}, in the decay (2)  the
cusp phenomenon also  takes place.\\
The above   expressions  completely solve the problem of generalization
of Fermi---Watson theorem to the case of two coupled channels with
different masses in the final state. To estimate the numerical difference of
proposed approach from usually  accepted,  we use the fact that  at the leading
order in chiral perturbation theory~\cite{Gass, ku} the isospin breaking change
 the relations (5) in the following  way:
\ba
a_n(s)&=&\frac{a_0(s)+2a_2(s)}{3}(1-\eta); a_c(s)=\frac{2a_0(s)+a_2(s)}{3}(1+\eta);\nn
a_x(s)&=&\frac{\sqrt{2}}{3}(a_0(s)-a_2(s))(1+\frac{\eta}{3});\eta=\frac{m_c^2-m_0^2}{m_c^2}
\ea
The  relations  (6)  are valid   also in the case of  isospin symmetry  breaking,
with simple replacement~\cite{Gass}  $k\to k_c$.
Substituting  these relations in the expression (12)  for phase shift relevant  to  decay (1) we obtain:
\ba
\delta_c &=&\arctan \left ( \frac{A\tan\delta_0^0+B\tan\delta_2^0}{1+\lambda W}\right ) +
\arctan\left (C\tan\delta_0^0+D\tan\delta_2^0\right)\nn
A&=&\frac{2(1+\eta)+\lambda(1-\eta)}{3}; B=\frac{(1+\eta)+2\lambda(1-\eta)}{3}; \nn
C&=&\frac{4\eta\lambda}{9}; D=-\frac{(3-\eta)\lambda}{3}; \lambda =\frac{k_n}{k_c};\nn
W&=&\frac{2}{9}(1+\frac{\eta}{3})^2(\tan\delta_0^0-\tan\delta_2^0)^2-\frac{1}{9}
(1-\eta^2)(2\tan\delta_0^0+\tan\delta_2^0)(2\tan\delta_2^0+\tan\delta_0^0)\nn
\ea
In  the fig.1 we depicted  the dependence  of  phase $\delta_c(s)$ obtained by expression (16) (solid line)
  and in the case  of exact isospin symmetry  (dotted  curve). The phases $\delta_0^0,\delta_2^0$ were calculated
  according to the  Appendix D of the review ~\cite{ACGL} for the values
of scattering lengths  $ a_0^0=0.225{m_c}^{-1}, a_0^2=-0.03706{m_c}^{-1}$. The isospin breaking effect 
provided by different masses  of neutral and charge pions    increase
 the s-wave  phase  and hence would has  impact on the value of scattering lengths
 extracted from $K_{e4}$ decays.\\
In the Fig.2  we  show  the invariant mass dependence of ratios
\ba
R_n=\frac{\tan\delta_n(s)}{k_na_0(s)}; R_c=\frac{\tan\delta_c(s)}{k_ca_0(s)}
\ea
The proposed  approach allows one to extract  from  decays  (1),(2)  besides the scattering length
$a_0^0$   the scattering length $a_0^2$, the challenge  which is absent in common approach.
 At present the high quality  data on $K_{e4}$ from  NA48/2 experiment at CERN are
published~\cite{bb}  and their  fitting by the expressions of present work would
be very useful and can shed light not only on the true values of
scattering lengths $a_0^0,a_0^2$, but also help  to understand the
limits and validity of proposed approach.\\
It is a pleasure to  thank  V.D. Kekelidze and D.T. Madigozhin for
useful discussions and permanent support.We are grateful to J.Gasser and A.Rusetsky
 for important comments and friendly criticism, which assist   to  improve the present
work and our understanding of considered above problems.

\newpage
\begin{figure}[ht]
\begin{center}
\includegraphics[scale=1.0]{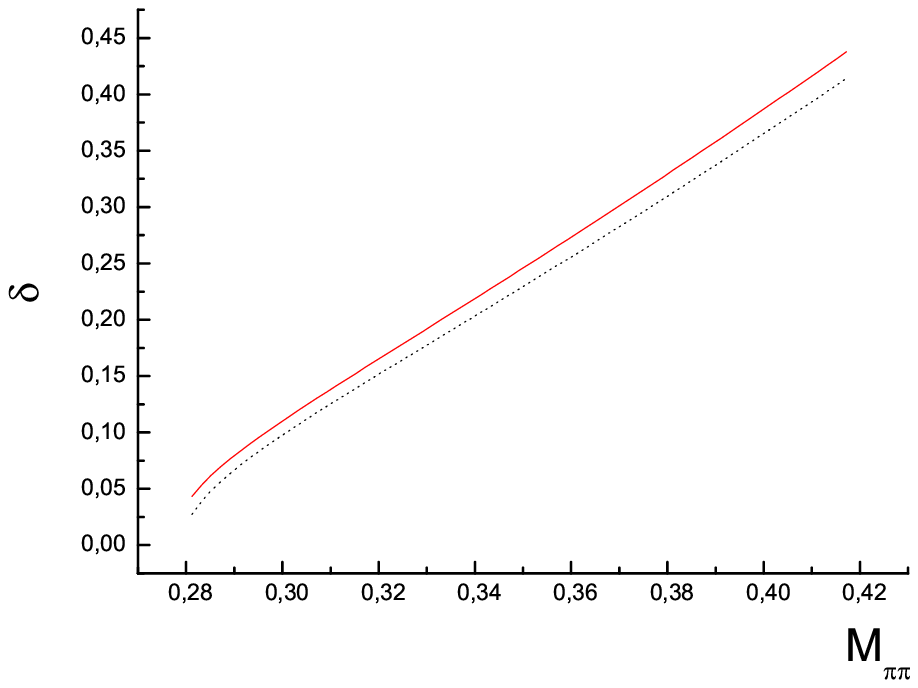}
\caption{The dependence of  s-wave  phase  $\delta_c(s)$  on invariant mass of  $\pi^+\pi^-$ pair
in the case of isospin symmetry  (dotted   line) and  isospin breaking case (solid line).}
\end{center}
\end{figure}
\newpage
    
\begin{figure}[ht]
\begin{center}
\includegraphics[angle=-90,scale=0.6]{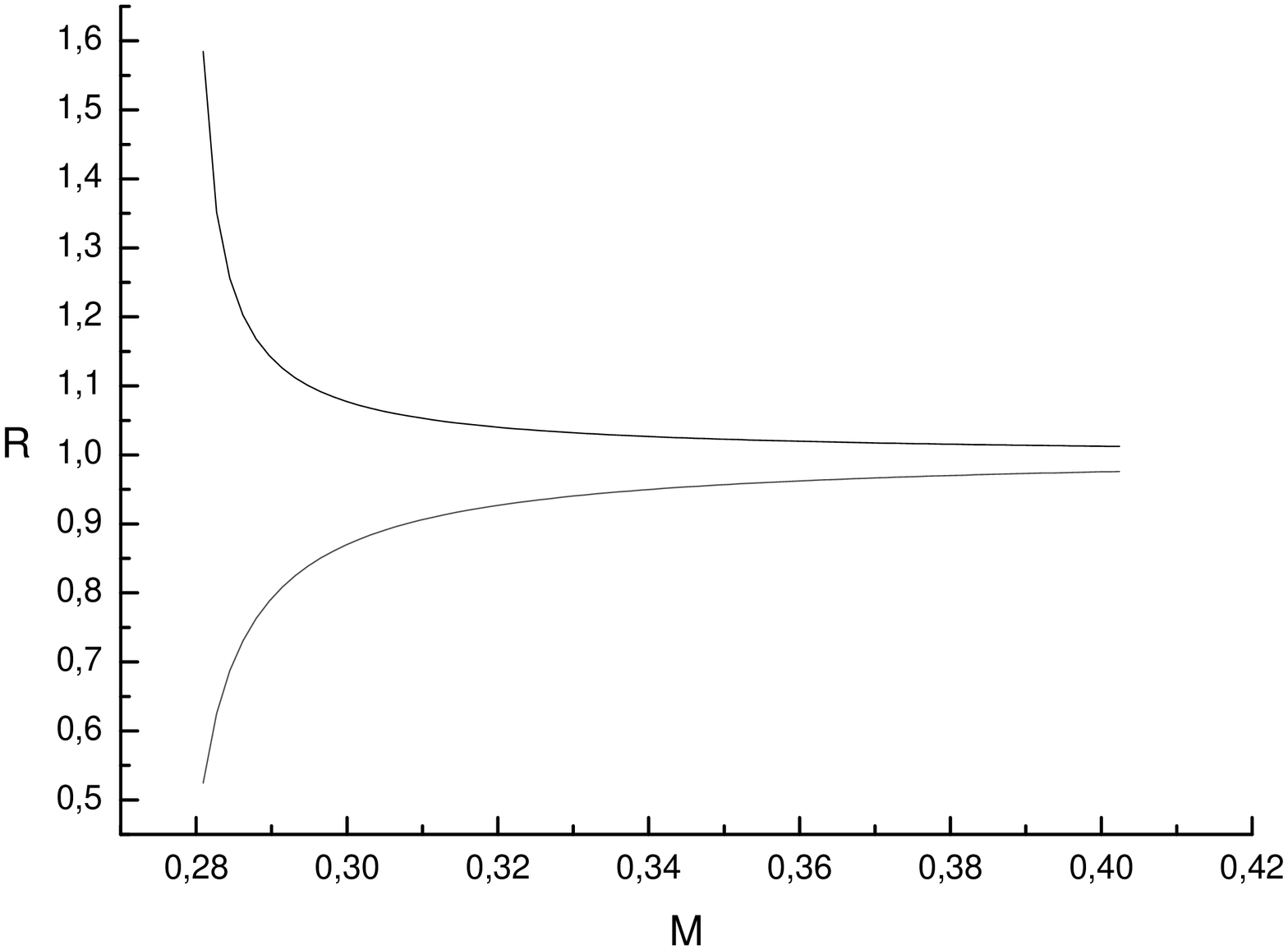}
\caption{The dependence of  s-wave  phase shifts ratios (17)  for  $\pi^+\pi^-$
(upper curve) and $\pi^0\pi^0$ (lower curve) on invariant mass of
the pion pair.}
\end{center}
\end{figure}

\end{document}